\documentclass[fleqn,usenatbib]{mnras}

\usepackage{newtxtext,newtxmath}
\usepackage[T1]{fontenc}

\DeclareRobustCommand{\VAN}[3]{#2}
\let\VANthebibliography\thebibliography
\def\thebibliography{\DeclareRobustCommand{\VAN}[3]{##3}\VANthebibliography}

\usepackage{graphicx}	% Including figure files
\usepackage{amsmath}	% Advanced maths commands
\usepackage{amssymb}	% Extra maths symbols
\newcommand{\be}{\begin{equation}}
\newcommand{\ee}{\end{equation}}

\newcommand{\lteff}{$\log{T_{\rm eff}}$}
\newcommand{\logg}{$\log g$}

\title[Short title, max. 45 characters]{MNRAS \LaTeXe\ template -- title goes here}

\title
{Simulating the outer layers of rapidly rotating stars}

\author[Robinson, Tanner \& Basu]
       {F.J. Robinson$^1$\thanks{Email: robinsonf3@sacredheart.edu},
            J. Tanner$^2$ and
S. Basu$^2$\\
$^1$Department of Chemistry and Physics, Sacred Heart University, Fairfield, CT, USA\\
$^2$Department of Astronomy, Yale University, New Haven, CT, USA\\}

\date{Accepted May 21, 2020. Received May 20, 2020; in original form April 2, 2020}

\pubyear{2020}

\begin{document}
\label{firstpage}
\pagerange{\pageref{firstpage}--\pageref{lastpage}}
\maketitle

\begin{abstract}
 This paper presents the results of a set of radiative hydrodynamic (RHD) simulations of convection in the near-surface regions of a rapidly rotating star. The simulations use microphysics consistent with stellar models, and include the effects of realistic convection and radiative transfer. We find that the overall effect of rotation is to reduce the strength of turbulence. The combination of rotation and radiative cooling creates a zonal velocity profile in which the motion of fluid parcels near the surface is independent of rotation. Their motion is  controlled by the strong up and down flows generated by radiative cooling. The fluid parcels in the deeper layers, on the other hand,  are controlled by rotation.

\end{abstract}


%%%%%%%%%%%%%%%%%%%%%%%%%%%%%%%%%%%%%%%%%%%%%%%%%%

%%%%%%%%%%%%%%%%% BODY OF PAPER %%%%%%%%%%%%%%%%%%

\section{Introduction}
\label{sec:intro}

Convection is one of the two dominant forms of heat transport in stars, and is also a significant source of uncertainty in stellar models. These models are predominantly one-dimensional, and incorporating a three-dimensional (3D) phenomenon such as convection into them, requires drastic approximations, such as the mixing length approximation (MLT) \cite{bohm1958wasserstoffkonvektionszone}. While MLT is successful in reproducing the properties of stars in near-adiabatic regions of efficient convection, there is evidence from helioseismic and asteroseismic observations that these approximations do not model the near-surface layers of a star particularly well \citep{abbett1997solar,kim1996modeling}. Of much more concern is the fact that the free parameters in the approximation directly control the radii of models, and thus these models cannot predict stellar radii without introducing additional constraints. To overcome this constraint, there are increasing efforts to simulate convection in the outer regions of stars on different parts of the Hertzsprung-Russell (HR) diagram \citep{magic2013staggerB,trampedach2013grid} for different metallicity \citep{ magic2013stagger,tanner2013variation} in order understand near-surface convection, and also to use the results of the simulations to improve stellar models \citep{mosumgaard2018stellar,spada2018improved}. These studies of convection have generally been limited to non-rotating stars. In this work we explore how rapid rotation affects near-surface convection.

Stellar convection has a huge range of spatial and temporal scales making it impossible to resolve all the dynamical and thermal scales in a numerical model. Consequently, there have been two approaches to modeling convection in stars; (i) global models of all or a large part of a star or (ii) local models of the surface layers of a tiny part of the star.

%two dynamical regimes, global or local
%\begin{enumerate}
%\item
%Global simulations of the rotating convection zone, for a spherical wedge, 
%or full spherical shell, geometry.
%\item
%Small scale (local) Cartesian box simulations of the outer surface layers.
%As these boxes are very small compared to the stellar radius, 
%the turnover time of the resolved eddies are many orders of magnitude less than the 
%rotation period of the star. As a result, it is generally believed that rotation does not need to be included in these models.
%\end{enumerate}

Global models have been developed for the Sun to model differential rotation \citep{BrunS02, MieschM07}, meridional circulation \citep{FeatherstoneN15} and the near-surface shear layer \citep{MieschM11,  HottaH14}. More recently, these techniques have been applied to other types of stars \citep{palacios2007simulation,brun2017differential}. These models typically include the entire convection zone in a full spherical shell geometry \citep{MieschM06, BrunS11, NelsonN18}, solving a truncated form of the fully compressible system called the anelastic approximation.  In this approximation, acoustic waves are filtered out and a linearized form of the governing equations is solved (see Appendix). This approximation limits the simulations to regions of efficient convection, and is therefore not suitable to study the effects of inefficient convection that causes the greatest uncertainty in stellar models.  Other global models \citep{RobinsonF01, KapylaP10},  solve the fully compressible system in a wedge covering   60-120 degrees in latitude and  longitude centered about the equator.

The global models span about 6-8 scale heights, whereas the convection zone on encompasses about 19 pressure scale heights. This means that the  resolved eddies in the global simulations are thousands of times larger than solar granules observed at the surface of the Sun,and hence again not suitable for studying the effects of inefficient convection on stellar structure.  Furthermore, the numerical energy flux at the lower boundary is millions of times larger than the actual radiative flux at that depth \citep{BrandenburgA05}. In these models the equation of state is typically an ideal gas and  radiation is approximated  by a conduction layer \citep{BrunS02}.

Local models are designed to accurately represent energy transport in the thin convection-radiation transition layer at the top of the star. \citet{1985SoPh..100..209N} was among the first to use local RHD simulations to accurately account for realistic near-surface convective dynamics and radiative energy transport, but the technique has since been used by numerous authors \citep[e.g.][]{1989ApJ...336.1022C,2004IAUS..224..139F,2013A&A...557A..26M}. These type of simulations are 3D Radiation Hydrodynamic (3DRHD) simulations and, as the name suggests, includes the effect radiation has on the dynamics caused by convection. Typically, the depth of a simulation box is between 3 and 5 Mm (about 8 pressure scale heights). The simulation is driven by an imposed energy flux at the base of the box. In these `realistic' simulations, this energy flux is close to the radiative flux $\sigma {{\rm T}_{\rm eff}}^4$, whereas in the global simulation it is many times larger. The simulations  use  realistic equations of state $P(\rho, e)$,  where $P$, $\rho$ and $e$ are the pressure, density and internal energy of the gas, respectively.  They also include the thermodynamic effects of ionization zones and 
use full radiative transfer \citep{SteinR00, FreytagB12, TrampedachR13}.  The majority of 3DRHD  simulations have been of the
Sun and other solar-like stars \citep{RobinsonF03, BeeckB12, TannerJ16,chiavassa2018stagger,collet2007three}, but there have also been realistic simulations of Subgiants \citep{RobinsonF04}, M-Dwarfs \citep{LudwigH02, WedemeyerS13},Betelgeuse \citep{freytag2002spots}, F stars such as Procyon \citep{RobinsonF05} and others  \citep{KitiashviliI16}.

%These ``realistic'' stellar  surface simulations can provide a 
%variety of useful information for stellar modelling of  stars, such as:
%%\begin{itemize}
%%\item
%%(1) Improving the mixing length in the surface layers \citep{TannerJ16,SpadaF18,ArnettD18}.
%%\item
%(2) Improving models of tidal dissipation from stars \citep{PenevK09,PenevK12,PatelR19}
%%\item
%(3) A  test-bed for turbulent closures 
%used to write  higher  order moments in terms 
%of lower order moments thereby  reducing 
%the order of moments in stellar models \citep{KupkaF06, KupkaF17, CaiT18}.
%%\item
%(4)Providing  model atmospheres to determine stellar metallicities \citep{AsplundM09, CaffauE08, JorgensenA18}.
%%\end{itemize}

%\subsubsection{Rotating 3DRHD simulations}

While the global models must include  the effects of rotation, the local models of "realistic convection" typically do not. This is because for most stars being studied \citep{mathur2011granulation},
% \citet{magic2013stagger, mathur2011granulation}), 
 the convective timescales in the computational domain  are tiny compared to the rotation period of the star.  For example, solar granules have lifetimes of about 10 minutes, while the Sun takes 25 days to spin once.  The importance of rotation associated with such flow, can be estimated from the ratio of this turnover time to the rotation period, which for solar granules is  about $3\times 10^{-4}$. Under such conditions the Coriolis terms in the Navier-Stokes equations will be insignificant compared to the other momentum terms.
%The effect of the centrifugal force can be estimated by dividing $R   by the gravitational
%acceleration $g$ in the simulation box.
The importance of the centrifugal force can be estimated by $R\Omega^2/g$, where $R$, $\Omega$ and $g$, are the radius, external rotation rate and acceleration due to gravity at the top of the star. For the Sun, this parameter is $   \approx 10^{-5}$, so the centrifugal force can also be ignored in local models of the Sun.
 
If the star is spinning fast enough, granules should start to feel rotation. This paper described a set of simulations of rapidly spinning stars that have rotation periods that are about an order of magnitude greater than the eddy turnover time. As an example, for a simulation of  $\delta$-Scuti, the estimated turnover time for a granule is about 30 minutes, while the observed rotation period can be as short as 6 hours \citep{SolanoE97, MolendaJ09}.
%In this case the ratio of turnover time to rotation period is about 0.1 and
%$R_{\rm{\delta-S}}\Omega^2/g $ is about 0.5.
Under these conditions, rotation cannot be ignored.

%However,  if a star has a
%thin convection zone and rotates rapidly enough, then
%then rotational effects on granule-scale structures can become quite significant.
%The $f$-plane
%approach has not been used in stellar radiation hydrodynamics because the convective eddies
%in those simulations (i.e. granules)
%typically do not have long enough turnover times  to ``feel'' rotation.
%However,  if a star has a
%thin convection zone and rotates rapidly enough, then
%then rotational effects on granule-scale structures can become quite significant.

%\begin{figure}
\section{Model Description}
Our numerical code solves the fully compressible Navier-Stokes equations with radiative transfer.  Details of the non-rotating version are provided in \cite{TannerJ12, TannerJ16} (and references therein). 
The code is an updated version of the code used in \cite{RobinsonF03, RobinsonF04} and is based on the code written by \cite{kim1998hydrodynamic}. Here, we outline its basic ingredients and key features among recent improvements. 

A 3D model is characterized by its surface gravity, chemical composition, and effective temperature, although only the former two properties can be directly set in a simulation. To produce a 3D RHD simulation, we begin with a stratification extracted from a 1D stellar model.  The initial stratification is taken from a 1D stellar model computed using MLT which is then relaxed to a steady state that is consistent with the realistic convective dynamics and radiative transfer.  In this work, we use the 1D stellar evolution code 
YREC \citep{demarque2008yrec} to obtain the initial stratification, but this does not affect the final relaxed state of the simulation, which can be characterized by the chemical composition, surface gravity, and effective temperature.  As a result of the inadequacy of MLT, the relaxed state of a 3D RHD simulation is not consistent with the stratification from the initial MLT model, and consequently, the radiative flux may converge to a value that differs from the 1D model.  Since it is determined from the radiative flux, the effective temperature is a computed property of a relaxed simulation, and we cannot set this as an input constraint.  The simulations presented in this paper are for a $\delta$~Scuti-type star with  \logg\ and \lteff\ of 4.21 and about 3.81, respectively  (the Sun has values of 4.44 and 3.76).

Due to the very large range of scales in a stellar  convection zone (e.g. 19 pressure scale heights for the Sun), the computational domain that is used is typically a tiny box, or thin shell, located at the top of the convection zone. The box has a width equal to that of a few granules and is deep enough so that further increases in depth do not alter the flow structure, with the top of the box located at least 1 pressure scale height (${\rm H_P}$) above the photosphere.  In this region, ${\rm H_P}$ is about 500km. The depth of the domain is important because boundary effects can be reduced (though never eliminated) by increasing the vertical extent until quantities within 1 ${\rm H_P}$ of the upper and lower boundaries remain unchanged, if further increases are made. The vertical walls are periodic and the horizontal walls are  impenetrable (closed box). To help reduce large horizontal velocities
(and shocks) in the low density layers near  the top of the box,
a no-slip boundary condition is imposed at the top
of the box, while the bottom of the box is stress free.
By comparing different 3D radiation hydrodynamics  simulations,
\cite{KupkaF05} and \cite{KupkaF08},
showed that the  effect of the top boundary  on the
underlying convection is felt only within a pressure
scale height of the top.
The effect of no-slip versus slippy top 
was tested for simulations of the Sun  in \cite{RobinsonF03}.
It was found that the boundary condition at the top had little impact  on the dynamical or thermal 
structure of the superadiabatic layer. 
A more recent comparison between the CO5BOLD, MURaM \citep{VoglerA05}
and Stagger codes \citep{NordlundA95}  was
made by \cite{BeeckB12}.  Again, even though the codes use different numerical schemes and
the runs are for different resolution, different top boundary conditions and different domain sizes, the dynamical and thermal structure below the photosphere look almost the same.
%To ease the time step restriction and keep full compressibility,
%we  use an implicit numerical
%scheme for the relaxation part of the simulation and then compute the statistics
%using a parallel explicit method e.g. \citep{ChanK07}.
%The model that we will use   is
%the  fully-compressible
%Chan-Kim-Sofia model (CKS)  which uses an implicit numerical scheme, ADISM (Alternating Direction Implicit
%on a Staggered Mesh) developed by \cite{ChanK82}.
%Given that one of the eventual aims of the study is to apply the results of rotating 3DRHD 
%models  to 1D stellar
%models, which are almost invariably calculated assuming gray atmospheres, we use gray
%radiative transfer calculations in our simulations.  

In the deeper, more opaque regions (optical depth of $\tau  > 10^4$), radiative transfer
is modeled by the diffusion approximation.
However, in the shallower regions, such as the superadiabatic layer,
photon mean free paths are
not small enough to use the diffusion approximation.
Here $Q_{rad}$, the energy transferred by radiation,
is computed as $Q_{\rm rad} = 4\kappa \rho (J $-$ B)$
where $\kappa$  is the Rosseland mean opacity, and the mean intensity
$J$ is computed by using the generalized three-dimensional
Eddington approximation of \cite{UnnoW66}. This formulation is exact for isotropic radiation in a gray atmosphere, and without requiring local thermodynamic equilibrium, the Eddington approximation describes the optically thick and thin regions exactly \citep{RuttenR95}.
A detailed comparison between using the 3D Eddington and  ray integration for radiation  has been done by  \cite{TannerJ12} who concluded  that as far as convection below the surface is concerned, there are only minor  differences in the dynamics or thermal structure between simulations using ray integration or 3D Eddington.
%Solar simulations with our 
%current formulation compare very well with other codes that use opacity binning 
%and integral over rays. Computationally, to solve the full radiative transport 
%equations is formidable, and typically only a few ray directions are 
%currently used in this approach (e.g., \cite{BeeckB12}). 
%Since ray integrals and opacity binning produce the same result as the 
%3D Eddington gray treatment, we are confident 
%that using 3D Eddington will produce accurate results.

The code is numerically stable for stars with surface gravity similar to that of the Sun or higher surface gravities, as well as somewhat hotter stars like F-type stars; no additional artificial viscosity is needed to stabilize the code.
The subgrid-scale model employed is the one from
\cite{SmagorinskyJ63}.

%\section{Rapid Rotation}
%In general, radiation hydrodynamical 
% simulations of the surface layers 
%do not  include rotation.
%This is because the 
%eddy turnover times of the convective motions in the surface layers are usually much less than the rotation period 
%of the star. For example, for the Sun the turnover time $\tau$ of granules is around 
%10 minutes, while the rotation period, $P$, is 25 days at the equator. 
%This gives a Coriolis number, $ Co = \frac{\tau}{P}$ of   
%0.0003, which means the Coriolis terms in the 
%Navier-Stokes equations will be insignificant compared to the other momentum terms. 
%The effect of the centrifugal force  can be estimated by dividing by the graviational 
%acceleration $g$ in the simulation box.
%For the Sun, $R_{\odot}\Omega^2/g \approx 10^{-5}$, so the centrifugal force can also 
%be ignored in RHD simulations.
%%  
%The type of stars for which our analysis is relevant are rapidly spinning stars that have 
%rotation periods within a factor of 10 of the eddy turnover time.
%For example, for the $\delta$-Scuti model in this paper, the estimated turnover
%time for a granule (based on a simulation)https://www.overleaf.com/project/5e32ffbdcba4c70001e2b37e  is about 30 minutes and 
%the rotation period can be as short as 6 hours \citep{SolanoE97, MolendaJ09}. 
%In this case the Coriolis number  is about 0.2 and 
%$R_{\rm{\delta-S}}\Omega^2/g $ is about 0.5.  
%Under these conditions, rotation cannot be ignored.    

\subsection{Adding rotation: the $f$ plane box}
\begin{figure}
  \centerline{\includegraphics[width=6cm]{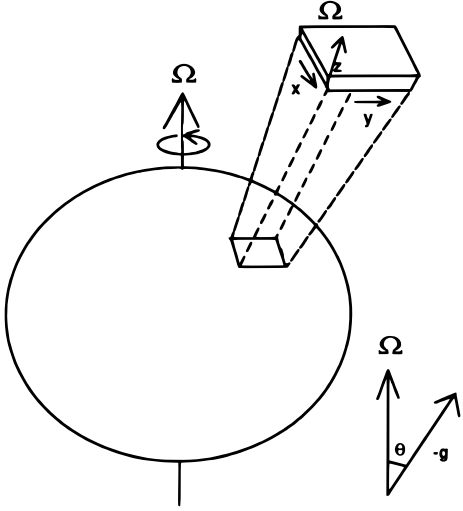}}
  \caption{The $f$-plane geometry.}
  \label{f-plane-geometry}
\end{figure}

To include rotation,  we 
employ the $f$-plane approximation \citep{pedlosky1988geophysical}. In our configuration, $f$ is defined as $2\Omega {\rm sin}\theta$, where $\theta$ is co-latitude. The geometry is shown in Fig. \ref{f-plane-geometry}.The computational domain  
can be thought of as a tiny part of a 
spherical shell at a particular  co-latitude.
The assumption of constant $f$  is good, provided the box size is small compared to the radius of the star, which is indeed the case of our simulations.
In the right handed Cartesian geometry
that we use, the $x$ direction
is towards the equator,  $y$  is in the zonal direction and the $z$ direction
is radially outwards. This means gravity $g$  is in the negative $z$ direction).
%In terms of a shell geometry,  $x$ is along 
%north-south (meridional) and $y$ is from west to east, i.e., along the zonal direction. 

%Laminar convection simulations in an $f$-plane box showed that a tilting of
%convection cells by rotation could generate Reynolds stresses that
%feedback on the mean flow \cite{HathawayD83}. 
%Numerical
%applications of  the $f$-plane  rotation version of our code to idealised polytropes  are described in
%\cite{ChanK01, ChanK07}.
The $f$-plane  configuration  has been used to simulate rotating turbulent convection
\citep{BrummellN96, ChanK01, ChanK07}, 
%penetrative
%incompressible convection \citep{TobiasS08} 
and convectively generated vortices  in Jovian planets \cite{ChanK13}.
All of these $f$-plane simulations are for idealised polytropic stratifications with input fluxes many orders of magnitude greater than the solar flux. 
The current application is different in that it is of realistic surface convection which contains both shallow and deep convection. 
\subsection{Non-dimensional units}
Physical quantities are scaled by their value at a reference level
located near the top of the box.   For instance, velocity is scaled
{\bf by } the isothermal sound speed, $c_{\rm s}=(p_{\rm top}/\rho_{\rm top})^{1/2}$,
and time by $d/c_{\rm s}$, where $d$ is the depth of the box.
In such units, the external angular velocity, $\Omega$,
is scaled by $c_{\rm s}/d$.

\subsection{Centrifugal and Coriolis Forces}
Results presented in this paper are for an $f$-plane box located at 
the equator ($\theta = \pi/2$,  $f=2\Omega$). In this configuration, the centrifugal force
is in the opposite direction to gravity and can be included by simply subtracting 
$R\Omega^2$ (where $R$ is the radius of the star) from the uniform gravitational acceleration.
As ${\bf \Omega}$ points along the $-x$ direction, there will be   
two non-zero components of the Coriolis force, $-(2\Omega v_z)\hat{y}$ and $(2\Omega v_y)\hat{z}$, 
where $v_y$, $v_z$, $\hat{y}$ and $\hat{z}$  are the zonal and vertical velocities and unit vectors in the 
$y$ and $z$ direction, respectively.   %During thermal relaxation the zonal 
%component of the Coriolis force can sometimes create a spurious 
 %net flow in the zonal direction. As there should be zero net flow between the fluid in the box 
%and the reference frame box, this is removed during relaxation by imposing the condition 
%$\langle \overline{\rho v_i} \rangle$ = 0 (i=,x,y).  

%\be
%2\bf{\Omega} \times \bf{V} = -\bf{y} (2\Omega V_z) + \bf{z}(2\Omega V_y)
%\ee
%The component of the Coriolis force in the zonal ($y$) direction will deflect coherent 
%vertical flow (i.e. plumes).
%The non-dimensional rotation rate
%is $\Omega=\Omega^* \times \tau_{sim}$  where $\Omega^*$ is the angular
%velocity in $s^{-1}$ and
%$\tau_{sim}$ is 515 seconds.
%The acceleration due to gravity is $g$ = 16044.3 $cm/s^2$
%The period of rotation, T,  is computed as $2\pi\times \tau_{sim}/\Omega$ .
%The second  simulation ($\Omega$ = 0.064) does not 
%take into  account the centrifugal force. The radius of the star is $R=1.055 \times10^{11}$ cm. Other quantities are described inthe ttext.
%

\subsection{Turbulent fluid properties}
Following the notation employed by \cite{ChanK01}, 
any turbulent fluid  quantity $q$ can be split into a mean and a fluctuating part,
\be
q = \overline{q}(z)+ q'(x,y,z,t).
\ee
The overbar represents a  combined horizontal and temporal average, i.e.
\be
\overline{q}(z) = \frac{1}{t_2-t_1} \int\limits_{t_1}^{t_2} \left(
 \frac{1}{
 (L_x L_y)
}
\int q   dx dy   \right)dt.
\ee
$t_1$, is a time  after the system has reached a self-consistent
thermal equilibrium (the thermal adjustment time). $L_x$ and $L_y$ are the
horizontal widths of the box in the $x$ and $y$ direction respectively.
The time required for statistical convergence is $t_2-t_1$.

The root mean square (r.m.s.) value of a quantity $q$ is defined as
\be
q''=\overline{q^2}-{\overline{q}}^2,
\ee
while the   correlation coefficient of two quantities $q_1$ and $q_2$,
is defined as
\be
\label{autoc}
 C[q_1'q_2']=\frac{\overline{q_1 q_2}-\overline{q_1}\hspace{1mm}\overline{q_2}}{q_1''q_2''}     
\ee

\begin{figure}
  \centerline{\includegraphics[width=9.5cm]{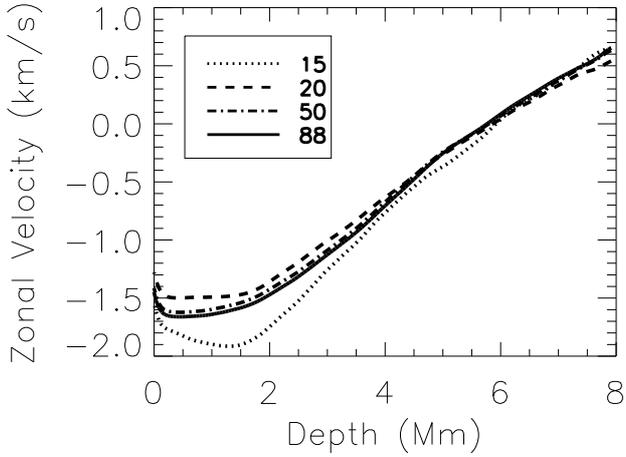}}
  \caption{Convergence of mean zonal velocity, ${\overline v_y}$. The plot  is  for case D.   Each line represents a different averaging times (in non-dimensional time units). ${\overline v_y}$ is converged after about 50 time units.}
  \label{converg_avymn_lnp}
\end{figure}

The symbol $\langle q \rangle$ denotes averaging over the height, $d$, of the box:
\be
\langle q \rangle = \frac{1}{d} \int{q}
\ee
%The time required for statistical convergence depends on the particular quantity being averaged
%with second order moments taking longer to converge than first order quantities, such 
%as the  mean temperature. 
Statistical convergence has been verified by comparing each quantity at different averaging times. If averaging over an additional 10 turnover times, did not change the value of a quantity by more than 1 \%,  then it was considered converged. For example, the mean zonal velocity in the box, $\overline{v_y}$ in case D is shown in Fig. \ref{converg_avymn_lnp}.  The zonal velocity, $v_y = v_{\phi} - R\Omega$, where $v_{\phi}$ is  the equatorial tangential velocity at a given point in the star.  It has been averaged over 15, 20, 50 and 88 time units. For rotating deep efficient convection,
%in an $f$-plane box
%placed at the equator, 
the slope of $\overline{v_y}$  should be $-2\Omega$ \citep{ChanK01}. As the deep layers take longer to converge (due to the longer turnover times), once the slope of $\overline{v_y}$ versus depth is constant near the base, one can assume $\overline{v_y}$ is close to convergence in the rest of the box.

%The acceleration due to gravity is $g$ = 16044.3 $cm/s^2$
%The period of rotation, T,  is computed as $2\pi\times \tau_{sim}/\Omega$ .
%The second  simulation ($\Omega$ = 0.064) does not
%take into  account the centrifugal force. The radius of the star is $R=1.055 \times10^{11}$ cm. Other quantities are described inthe ttext.
%

%if omega(non-dim) = 0.1 then omega(dim)=0.1/500 = 0.0002 /s 
%Period = 2pi/(omega(dim)) = 2pi/(0.0002) = 10 hrs)

\section{Results}
\label{prelim-results}

%Table~\ref{models} shows some characteristics of the five simulations performed in this study. 
Table ~\ref{table:models} lists some properties  of the simulations. 
Columns 1-6  are model identifier,  rotational velocity $v_{\rm rot} = R \Omega$, non-dimensional angular velocity ${\Omega}^{*}$,
relative reduction in $g$
due to centrifugal force, r.m.s. velocity fluctuation 
and  r.m.s. velocity, respectively.
The r.m.s. velocity fluctuation is  computed as the depth average of $v''$,
where  $v''=({{v_x}''}^2+{{v_y}''}^2+{{v_z}''}^2)^{1/2}$. The r.m.s. velocity is 
computed similarly, but without subtracting the mean.  

From $v''$, various non-dimensional turbulence parameters are computed and presented in 
columns 7-9. 
The Coriolis number (inverse Rossby number), Co, is defined 
as ${\rm Co}=\Omega d/\langle v'' \rangle$. The Reynolds numbers, Re,  which compares the magnitudes 
of the inertia and viscous terms, is defined as 
$\langle v'' \rangle d/\langle \overline{\mu}/\overline{\rho}\rangle$ (where 
$\mu$ and $\rho$ are the subgrid scale  dynamic viscosity and the density). 
The Taylor number Ta, which  compares the Coriolis and viscous terms, 
is computed as $(2\Omega d^2/\langle \overline{\mu}/\overline{\rho}\rangle)^2$. 
 
\begin{table*}
\begin{minipage}{140mm}
\caption[Grid refinement study]{Dynamical Characteristics of Simulations}
\begin{tabular}{ccccccccc}
\hline
Model & $v_{\rm rot}$ (km/s)&  $\Omega^*$  &$ (g-R{\Omega}^2)/g$
& $\langle v'' \rangle  ({\rm km/s})$
& ${\langle {\overline v^2} \rangle}^{1/2}  ({\rm km/s})$ &
Co & Re & Ta \\
\hline
\\
A & 0  & 0.0   & 1.0  &   4.72 &  4.86 & 0.0    &2070   & 0.0\\
B & 153        & 0.064 & 1.0  &  4.38 & 4.62 & 0.21   &1760   & $ 5.5\times{10^5}$\\
C & 153        & 0.064 & 0.9  &  4.31 &  4.51 & 0.21   &1810   & $ 6\times{10^5}$\\
D & 184        & 0.089 & 0.81 &  4.20 &  4.47 & 0.31   &1865   & $ 1.3\times{10^6}$\\
E & 307         & 0.128 & 0.61 &  3.98 &  4.42 & 0.46   &1893   & $ 3.1\times{10^6}$\\
\hline
\end{tabular}
\label{table:models}
\end{minipage}
\end{table*}

%Following the notation in \cite{ChanK03}, the Coriolis number, $Co$, 
%is defined as $\Omega \tau_{\rm t}$  where $\tau{\rm t}$ is the local turnover time.
%To characterise the overal effect, the average $Co$ is computed as  $\Omega d/<v''>$ 
%(equal to the inverse Rossby number, $Ro$).
%So $Ro$ ranges from $\infty$ to 2. 
%The acceleration due to gravity is $g$ = 16044.3 $cm/s^2$
%The period of rotation, T,  is computed as $2\pi\times \tau_{sim}/\Omega$ .
%The second  simulation ($\Omega$ = 0.064) does not
%take into  account the centrifugal force. The radius of the star is $R=1.055 \times10^{11}$ cm. Other quantities are described inthe ttext.

\subsection{Thermal structure}

One measure of convective instability is the superadiabaticity, defined as $\nabla -\nabla_{\rm ad}$ where $\nabla= {\rm d ln T/ d  ln P}$ and $\nabla_{\rm ad}$ is the adiabatic temperature gradient. The total pressure is the sum of the gas, P and turbulent pressure ${\rm P_{turb}} = \overline {\rho} (v_z'')^2$.
For deep efficient convection, $\nabla - \nabla_{\rm ad}$ is slightly above zero ($\sim 10^{-6}$ in the solar  convection zone), while for shallow inefficient convection it is  of order unity. %  lower the value of $\nabla -\nabla_{\rm ad}$, the more efficient the convection.  

\begin{figure}
  \centerline{ \includegraphics[width=10cm]{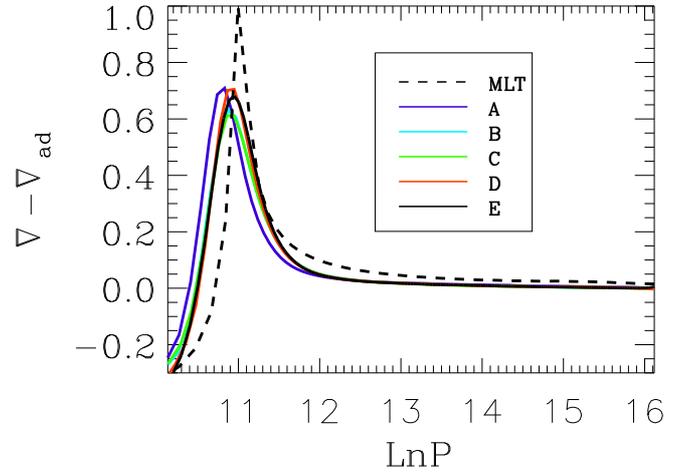}}
  \caption{Superadiabaticity versus LnP  for the original stellar model (MLT) and simulations A, B, C, D and E.}
  \label{sal}
\end{figure}

Fig. \ref{sal} shows $\nabla -\nabla_{\rm ad}$, plotted against LnP for the  stellar model (using MLT) that sets the initial conditions for the 5 simulations described in Table \ref{table:models}. The superadiabaticity from 3D RHD simulations is significantly different than that of the MLT model. This difference is  due to the realistic convective and radiative transport near the surface.  

\begin{figure}
  \centerline{ \includegraphics[width=10cm]{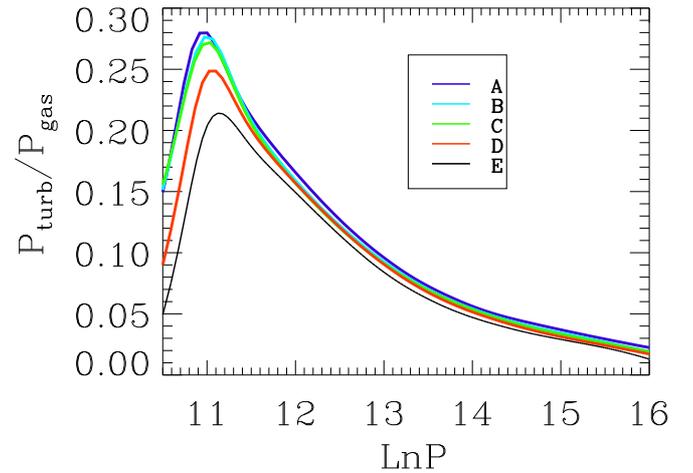}}
  \caption{Turbulent pressure divided by gas pressure versus LnP  for  simulations A, B, C, D and E.}
  \label{pturb}
\end{figure}

Another characteristic that the MLT approximation of convection does not account for is turbulent pressure.  The turbulent pressure as a fraction of by the gas pressure is plotted in Fig. \ref{pturb} for each of the simulations, while the turbulent pressure for the MLT model is not shown because it is zero.  

Comparing the MLT model with simulation A in figure \ref{sal}, one can see that the turbulent pressure has pushed the superadiabatic layer (SAL) outwards by about  0.5${\rm H}_{\rm P}$ from its original position. The photospheric surface is moved out by a similar amount.
Adding rotation reduces the turbulent pressure, so that the SAL is not pushed out quite as far. Simulations D and E have slightly higher SAL peaks than B and C, signifying slightly less efficient convection.  
%B and C have the the same peak superadiabaticity and turbulent pressure. Models B and C have the same rotation, but different centrifugal forces. This suggests that the difference between D and E is due to the difference in Coriolis force, rather then the change in centrifugal force. 

%The centrifugal force has a much larger impact on the
%turbulent pressure than the Coriolis force. This is principally due to the
%change in  $g$. Reducing $g$  and resulting change in the gas pressure. 

\subsection{Plumes}
%NOT SURE IF WE WANT GRANULES? 
%.A horizontal slice of the temperature filed of the convection box, Fig \ref{granules}, shows the broad upwelling regions (light color) surrounded by cool downflows (dark color).   
Fig. 5 % \ref{fig:y-z plumes}
shows an instantaneous vertical cross-section in the $y$-$z$  plane of temperature from simulations A, B and E.  The cooler coherent vertical structures (shaded) extending from the surface of the star to lower heights are called plumes. They are generated  by radiative cooling at the photosphere. As the rotation is increased the vertical extent of the plumes reduces. The effect is visible in the figure, where plumes in the non-rotating simulation (top panel) extend down to at least -3Mm below the surface, while for the rotating cases (lower panels), they only reach a depth of about -2Mm. The vertical correlation or mixing length are significantly reduced by rotation. 
%The Coriolis force. $F_c$  is defined as 
%\be
%{\bf F_c} = -2 {\bf{\Omega}} \times {\bf{v}} = -2 \Omega v_z \hat{y} + 2\Omega v_y \hat{z}
%\ee 
%Plumes  have  $v_z < 0$ and will be deflected in the $y$ direction. 

\begin{figure}
    \centerline{\includegraphics[width=\columnwidth]{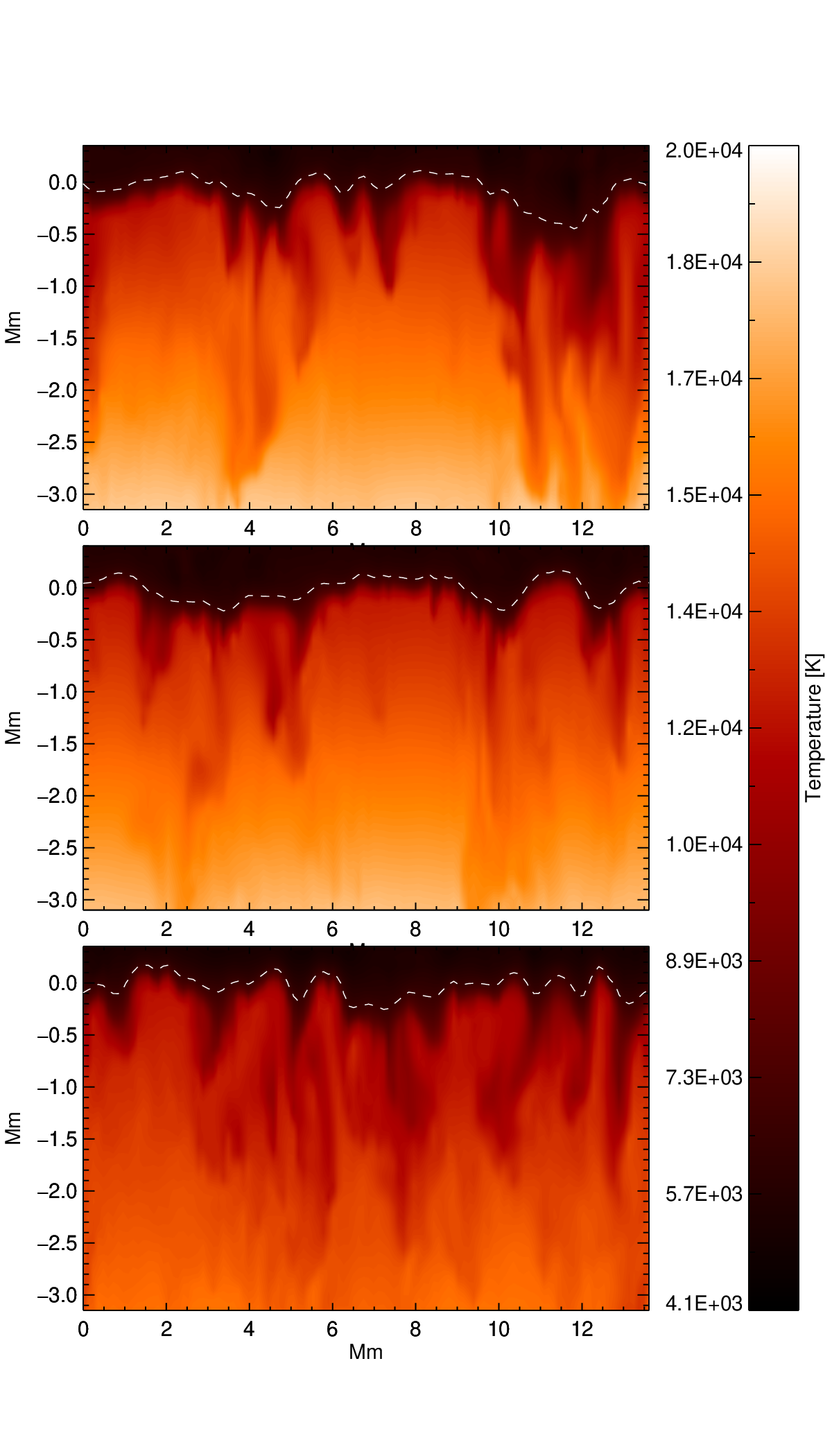}}
    \label{fig:y-z plumes}
    \caption{Snapshots of the upper part of a vertical cross-section of temperature field for $\Omega^* = 0$ (upper panel), 0.064 (middle panel) and 0.128 (lower panel). These are taken from models A, B and E. Dashed lines shows the $T=T_{\rm eff}$ surface across each slice, and depth is measured with respect to the mean $T_{\rm eff}$ surface. The principle effect of rotation is to reduce the penetration depth of the plumes.  
   }
\end{figure}

\subsubsection{Vertical scale of convective eddies}
\label{eddies}

While the visual inspection of instantaneous snapshots is informative, to better understand what is happening in the turbulent flow we need to look at statistics averaged over many turnover times. Although there is no `mixing length' in 3D RHD simulations, there are several ways of measuring an equivalent quantity.  
 
To estimate the length of a characteristic eddy, $l_z$, we computed the spatial auto-correlation (equation \ref{autoc}) of the $v_z'$ field, and define $l_z$ as the distance at which $C[v_z'v_z']$ falls to $0.5$. If we assume that the convective eddies have  an aspect ratio of unity, then $l_z$ gives us an idea of the size of the eddies in a turbulent fluid.

\begin{figure}
    \centerline{\includegraphics[width=10cm]{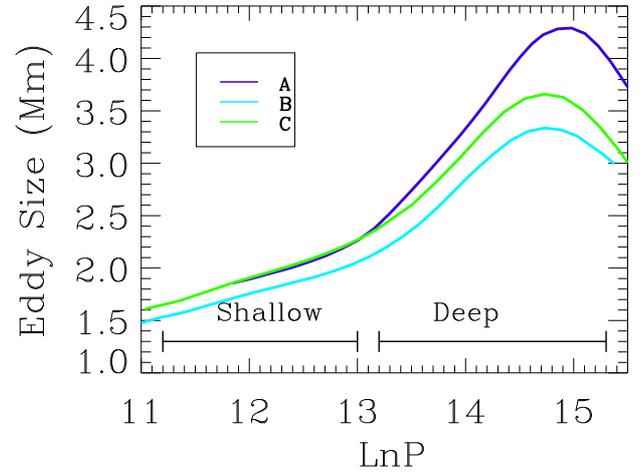}}
    \caption{Vertical eddy size, $l_z$,  versus LnP for models A, B and C.  Coriolis force does not appear to affect the near-surface eddy sizes, while the centrifugal force reduces the eddy size at all depths.  The effect of rotation is more pronounced in the deeper regions of efficient convection.}
    \label{fwhm}
\end{figure}

Fig. \ref{fwhm} shows $l_z$ for models A, B and C. Comparing Model A  and  model B, one can see that rotation has much more impact on the eddy size in the lower half of the domain. In the shallow regions, $l_z$ is  about 10 \% smaller for model B compared to model A, whereas in the deep layers, it is about  30\% smaller.  The effect of rotation on eddy size is three times greater for deep efficient convection, than it is for shallow inefficient convection.
The reduction in correlation length with depth is consistent with the decrease in penetration depth of the plumes in the previous figure.  
In general, the size of the eddies increases, until they are about one scale height from the base. The drop off in eddy size near LnP = 15 is due to the approaching  impenetrable bottom surface. 

Comparing simulations C and B illustrates the effect of the centrifugal force on the eddy size. These two simulations have the same Coriolis force, but $g$ is 10 \% lower in C relative to B to account for the centrifugal force. The reduction in $g$ between C and B, results in an increase of  $l_z$ of about 10 \%. This is because $l_z$ is proportional to the local scale height, ${\rm H_P}$ which is  $ \approx {\rm RT}/g$, where R and T are the universal gas constant and temperature, respectively \citep{RobinsonF04, magic2015stagger}. 

%So  decreasing $g$, increases  ($/approx RT/g$), which assuming a constant mixing length, will increase $l_z$
 
\begin{figure}
    \centerline{\includegraphics[width=10cm]{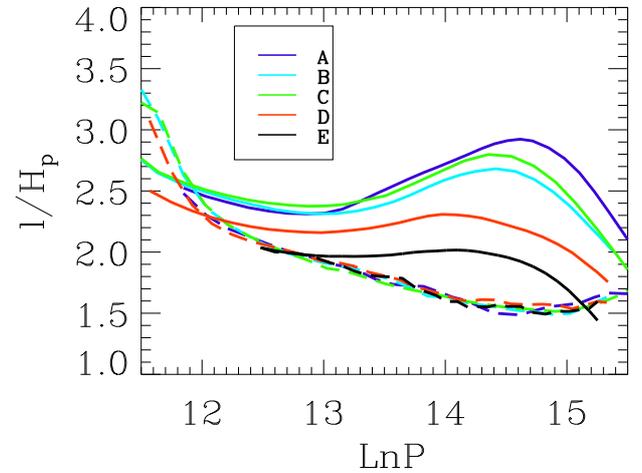}}
    \caption{Mass mixing length ($l=l_m$, long dashes) and eddy size ($l=l_z$, solid) divided by local pressure scale height. Rotation has a large effect on eddy sizes (especially in the deeper layers), but no effect on mass mixing length.}
    \label{alpha}
\end{figure}

Another convenient measure of the stellar mixing length is the so-called `mass-mixing length', $l_m$, described in \cite{trampedach2011}. Using their equation for $l_m$ we computed $l_m$ for the upflows as, 
\be
l_m (z)= 
\Bigl | {
 \frac{ {\rm dln}\overline{\rho} }  { \rm{dz} }
+\frac{ {\rm dln}\overline{v_z} }   { \rm{dz} }
+\frac{ {\rm dln}\overline{A}   }   { \rm{dz} }
}
\Bigr| ^{-1}
\ee

where A is the horizontal area occupied by upflows and overbars denote horizontal and time averages. Fig.~\ref{alpha} shows $l_z$ (solid lines) and $l_m$ (long dashed lines), each divided by ${\rm H_P}$, versus depth. While the $l_z$ depends on rotation, $l_m$ is almost the same for all five simulations. Rotation does not appear to have any impact on the mass mixing length.

% The vertical extent is reduced by the zonal component of the 
%Coriolis 
%force which deflects the path of downdrafts and thus reduces FWHM in the deeper regions
%%(compare A and B). 
%However, if the centrifugal force is included then $g$ is reduced, which increases the size of
%%the granules (FWHM $\propto g^{-1}$). 
%
%The correlation  length is proportional  to the 

%ressure scale height which is ${\rm P}/\rho g$ 

\subsubsection{Kinetic energy flux}
\begin{figure}
    \centerline{\includegraphics[width=10cm]{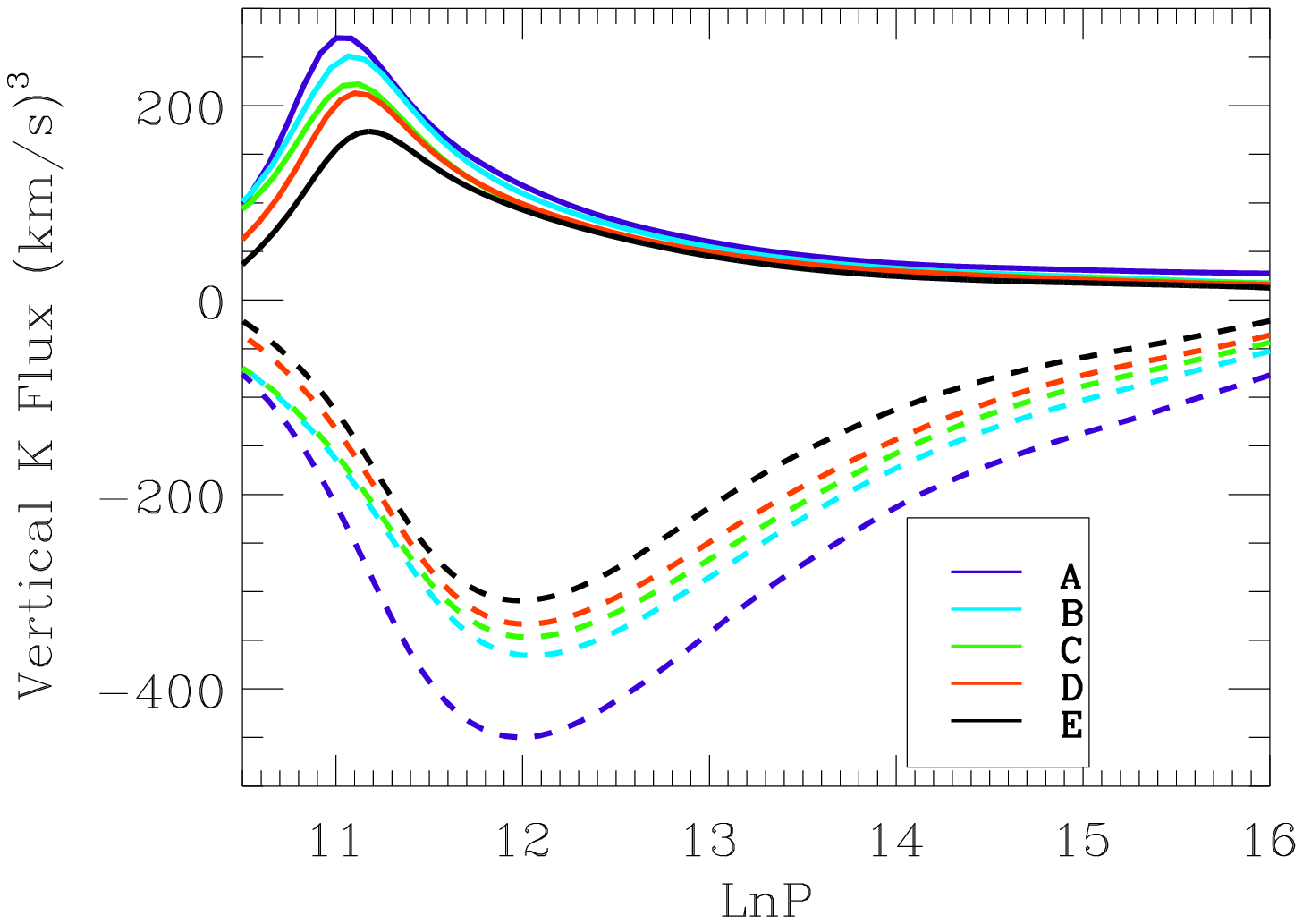}}
    \caption{Flux of kinetic energy per unit mass
         for models A, B, C, D and E. The updrafts and downdrafts are represented by solid and dashed lines, respectively}
    \label{keflux}
\end{figure}

Fig.~\ref{keflux} shows the flux of  kinetic energy  per unit mass by downdrafts (dashed)  and updrafts (solid) for simulations A-E. The quantity plotted is $\overline{v_z K}$ where $K=1/2({{v_x}^2+{v_y}^2+{v_z}^2})$. Comparing this plot with the plot of the SAL (Fig \ref{sal}), one can see that the flux of K by upflows is strongest near the peak of the SAL (LnP = 11) and by downflows is strongest near the base of the SAL  (LnP = 12). The flux by the downflows is about double that by the upflows. The effect of rotation on the flux of K is quite significant. The peak downflow and upflow of K drops by roughly 30\% between model A and E. 
%Rotation reduces the flux of K by about 50 \%. 
%It appears that the change in the flux only depends on the change in $\Omega$ not on the change in $g$. The effect of rotation is far more pronounced

\subsection{Effect of rotation on shallow vs. deep convection}

\begin{figure}
  \centerline{\includegraphics[width=10cm]{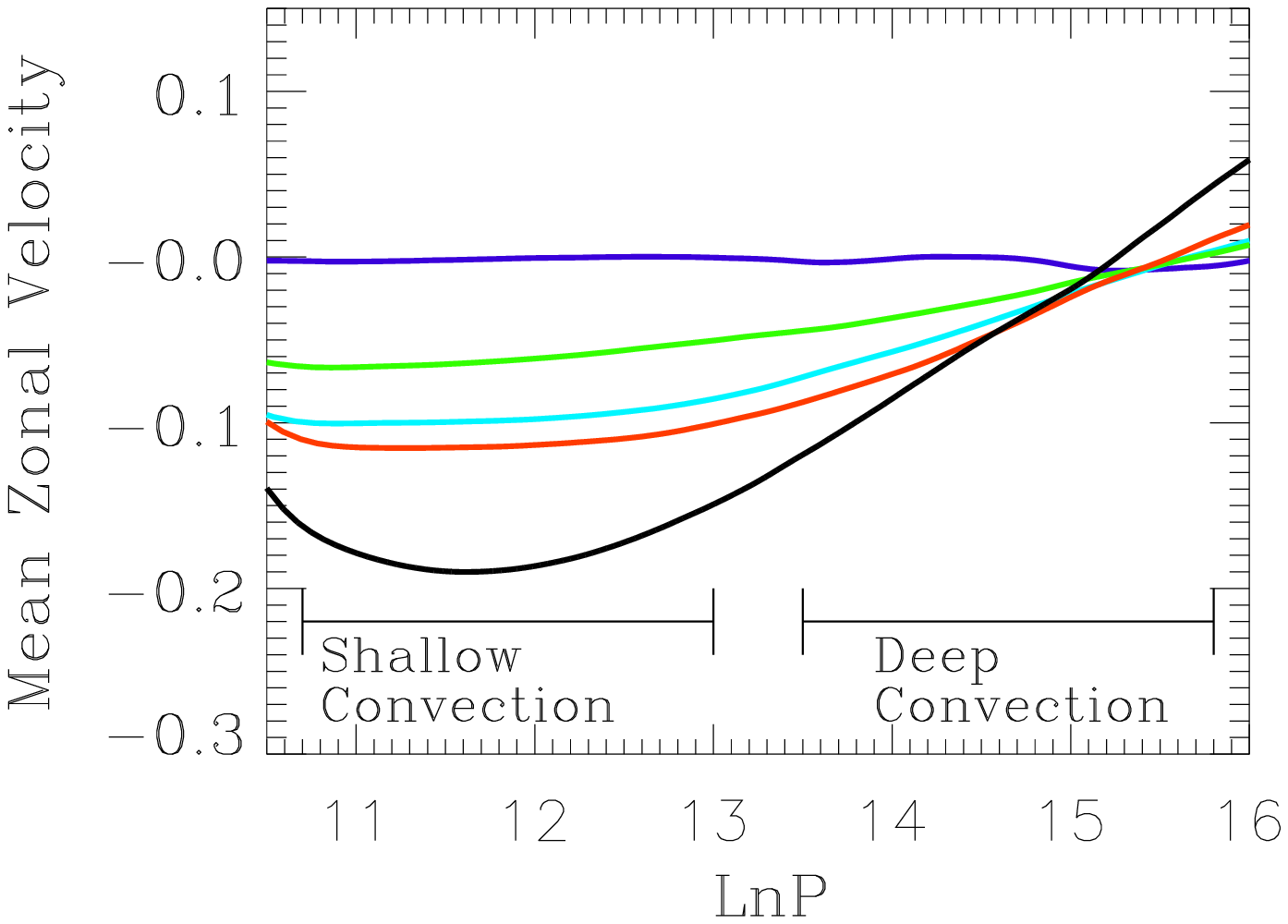}}
  \centerline{\includegraphics[width=10cm]{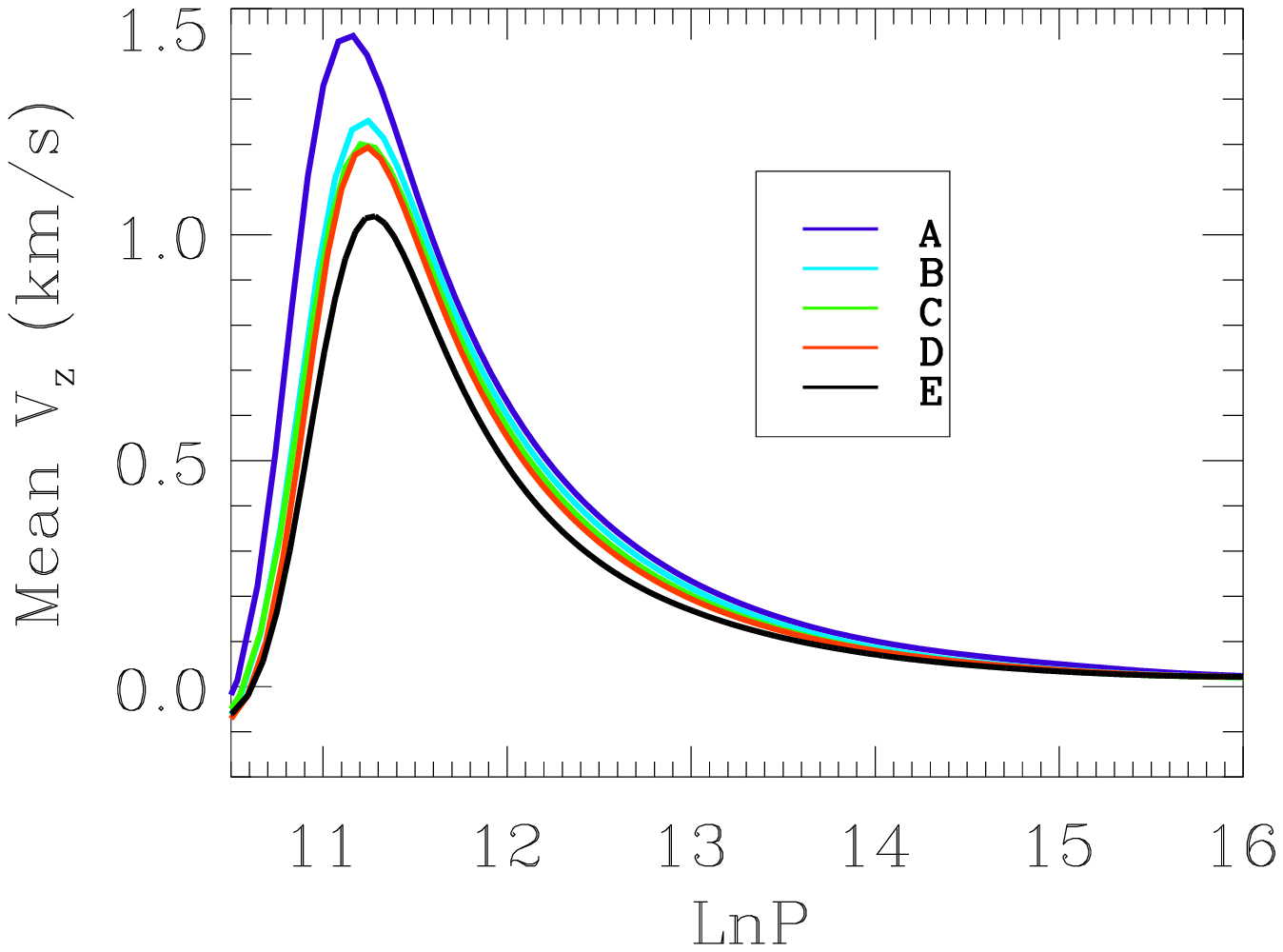}}
  \caption{Non-dimensional $\overline{v_y}$ (Upper panel) and $\overline{v_z}$  (lower panel) versus depth.}
  \label{meanvels}
\end{figure}

One of the robust features of efficient (deep) convection in $f$-planes at the equator is the linear increase in  mean zonal velocity, $\overline{v_y}$ with depth. In the study of deep convection by \cite{ChanK01}, the slope was shown to be $-2 \Omega^*$ regardless of input flux or rotation rate. The upper panel of  Fig.~\ref{meanvels} shows $\overline{v_y}$ versus depth for the five models. The closeness of the slope of $\overline{v_y}$ to $-2 \Omega^*$ in the deep layers indicates convergence. There is a clear distinction between deep and shallow convection. In the deep region the slope of $\overline{v_y}$ versus depth is constant (and equal to $-2 \Omega^*$), while in the shallow layers the zonal velocity is approximately constant.

%The sharp drop of $\overline{v_y}$  to zero at the top is due to the no-slip upper boundary condition.

In \cite{ChanK01} the $-2 \Omega^*$ shear was explained by conservation 
of angular momentum. We will briefly repeat their explanation. To conserve angular momentum, the angular velocity of fluid parcels moving outwards will decrease and that of parcels moving inwards will increase. Consider the effect of only the Coriolis force on a fluid parcel moving upwards or downwards at the equator  (where all quantities are in non-dimensional units):
\be
\frac{dv_y}{dt} \sim  -2\Omega v_z
\ee 
\be
\frac{\Delta{v_y}}{\Delta t} \sim -2 \Omega \frac{\Delta{z}}{\Delta t}
\ee
\be
\frac{\Delta{v_y}}{\Delta{z}} \sim -2 \Omega   
\ee

This explains why  $v_y$  decreases with a slope of -2$\Omega^*$ inwards for deep efficient convection. However, for shallow convection, we find that $v_y$ is roughly constant, so that
\be
\frac{\Delta{v_y}}{\Delta{z}} \sim  0
\ee

The change in slope signals the transition from shallow inefficient convection (granulation) to deep efficient convection.  Without the SAL, the ${\overline{v_y}}$ would have a constant -2$\Omega^*$ variation with depth throughout the domain.

The analysis in section \ref{eddies} showed that the eddy size is weakly dependent on  rotation in the shallow layers. It is restricted by the granulation. This is why for shallow convection, $l_z$ is similar for rotating and non-rotating convection. However, in the deeper layers $l_z$ is significantly reduced by rotation. In the deep layers fluid parcels are not controlled by the upflow/downflows associated with the SAL, they are free to feel the full effect of rotation.  The lower panel of Fig.~\ref{meanvels} shows the  mean vertical velocity as a function of depth. When the  mean vertical velocity is high,  the zonal velocity is constant. When the mean vertical velocity is small (below LnP = 13), the zonal velocity decreases inwards with a slope of -2 $\Omega^*$.
\subsection{Discussion}

Why don't the fluid parcels in the shallow layers appear to conserve angular momentum as they move radially inwards or outwards? The Coriolis force acts throughout the fluid, but in the shallow regions its effect on the vertically moving fluid parcels is much weaker than it is in the deeper layers.  The shallow layers, which are occupied by the granules,  move more like a rigid body with little variation with depth. This type of `rigid-body rotation' was suggested  by  \cite{foukal1975rotation}, as a feature of the near-surface shear layer in the Sun. They estimated  that neither the drag produced by viscosity nor the drag from magnetic flux tubes, were large enough to fix the rotation velocity in the upper layers/photosphere, and that some other mechanism was responsible. In the recent simulations of the near surface shear layer by \cite{matilsky2019role}, they describe `rotationally unconstrained' fluid as being associated with  fast down-flowing plumes.  In our simulations, these `rotationally unconstrained' fluid parcels are at the depths occupied by the strongest upflow and  downflows. 

%However, at the equator, the dominant terms in the horizontally averaged momentum balance are 
%Coriolis force and viscosity \citep{ChanK01}, so it is also possible that in our simulation, viscosity could play some role in sustaining  the uniform  zonal velocity in the shallow layers. Clearly, further analysis is required to precisely determine cause and effect.

\section{Summary and Conclusions}
Rotation is typically ignored in simulations of granulation in the outer layers of stars. This is a reasonable  assumption, provided the star does not spin too fast. If it does, then rotation weakens the turbulent vigor, which lowers the superadiabatic layer. In addition, even though rotation is the same  throughout the box,  its effect  is much weaker in  the shallow layers compared to the deeper layers. 
%To understand why, consider the effect of  rotation on a fluid parcel. 
To conserve angular momentum, the angular velocity of the parcel should increase  as it moves radially inwards towards the rotation axis (or equivalently to the bottom of the $f$-plane box). However, in our simulations we find  that the angular velocity is approximately constant over the first 2 scale heights below  layer (SAL) and then linearly increases inwards. Fluid parcels moving towards the rotation axis do not increase their angular  velocity  until they are below the SAL. We attribute this restriction on angular velocity to the strong vertical motions associated with the SAL. It appears that the short-scale faster overturning motions in the SAL are too quick to be significantly affected by the Coriolis force. 

One way to test this hypothesis, is to compare the timescales of fluid parcels in the shallow and deep layers, to the rotation period of the box. A rough estimate for the timescale of an eddy is $\tau_{\rm eddy} = l_z/{\overline v_z}$, where $l_z$ is the eddy size and ${\overline{v_z}}$ is the mean vertical velocity. Taking values for model B directly from Fig.~ \ref{fwhm} and the lower panel of Fig.~ \ref{meanvels}, we find that  in the shallow region, $l_z \sim$ 2000 km and ${\overline{v_z}} \sim 1$ km/s, implying $\tau_{\rm eddy} \sim$  30 minutes, which is 24 times smaller than the 12 hour rotation period. In the deep layers, $l_z \sim$ 3000km and ${\overline{v_z}} \sim$ 0.1 km/s, implying a timescale of about 8 hours, much closer  to the rotation period of the box. These estimates, though crude, support the  hypothesis that the external rotation rate of the box will impact  fluid parcels  in the deeper layers much more than parcels just below the surface of the star. This can also be expressed in terms of a non-dimensional flow parameter called the Coriolis number (or inverse Rossby number), Co =$\Omega \tau_{\rm eddy}$. Rotational effects are less prominent in the surface layers because Co is much smaller there. The variation of ${\overline{v_y}}$ with Co for a much larger range of Co values,  is described in \cite{chan2003numerical}

Results that may be of interest to stellar modellers, are the effect of rotation on the  SAL, the turbulent pressure and  the invariance of the mass mixing length to rotation.
The impact of rotation on convection near the top of the star could affect the  macro and micro turbulence parameters computed from 3D simulations \citep{SteffenM09}, and particularly in hot stars, such as F stars 
 with ultra thin convection zones, the change in macroturbulence could impact the interpretation of stellar spectra \citep{SaarS97}.
These results are also relevant to convection in the outer layers of young active stars which rotate much more rapidly \citep{BrownB08}.

The results described in this paper apply to realistic surface convection at the equator. We have not yet examined  different latitudes nor examined the  effect of a spherical shell geometry ---  since the $f$-plane box has periodic boundary conditions, it is unable to produce a realistic meridional circulation. However, out aim was to study the effect of rotation on convection, and not large scale flows, and our results show that there are measurable effects which can impact stellar structure,

%%\begin{figure}
%\begin{minipage}[b]{0.4\textwidth}
%    \includegraphics[width=9cm]{figures/Pturb_omega.eps}
%    \caption{Turbulent pressure vs. depth for rotation  rates, $\Omega$  of 0.0 and 0.1.}
%    \label{Pturb}
%  \end{minipage}%
%\hspace{2cm}
%\begin{minipage}[b]{0.4\textwidth}
%    \includegraphics[width=9cm]{figures/KEturb_omega_ds.eps}
%     \label{TKE}
%    \caption{Turbulent kinetic energy vs. depth  for rotation  rates, $\Omega$  of 0.0, 0.05  and 0.1.}
%  \end{minipage}
%\end{figure}

%\nobibliography{alldat}
\section*{Acknowledgments}
This work was supported in part by the facilities and staff of the Department of Astronomy in the Yale University Faculty of Arts and Sciences.

\bibliographystyle{mnras}
\bibliography{Rob1} % if your bibtex file is called example.bib

\section{Appendix: Compressibility}
As the time step in the numerical schemes is limited by the time for a sound wave to
cross between two grid points (known as the C.F.L. condition), to reach a  steady state, requires a huge number of timesteps. To ease this restriction,
many of the global models use ``sound proof'' equations such as the
Boussinesq \citep{spiegel1960boussinesq}  or anelastic \citep{gough1969anelastic} approximations,
in modelling solar differential rotation \citep{MieschM06,FeatherstoneN15}.
By comparing  the momentum equation for the
Boussinesq, anelastic and  fully compressible models,
one can see how baroclinicity is modelled in each system.
Excluding rotation and viscosity (to simplify the analysis), the momentum equation can be written as
\be
\rho \frac{D {\bf v}}{Dt}=-\nabla p' - g\rho'{\bf e_k}
\label{momentum}
\ee
where $p'$ and $\rho'$ are the  perturbation pressure and density (i.e. the hydrostatic part has been removed),
and the material derivative is
\be
\frac{D} {Dt}=\frac{\partial} {\partial t} + ({\bf v}\cdot \nabla).
\ee
Depending on the  particular choice of $\rho$ in the inertia term one obtains,
\begin{enumerate}
\item
the incompressible Boussinesq model: $\rho= \rho_{\rm o}$ : constant reference state
\item
the anelastic model: $\rho =  \rho_{\rm o}(r)$ : spherically symmetric reference state
\item
the fully compressible model: $\rho=\rho(r,\theta,\phi,t)$  : full baroclinicity
\end{enumerate}
For the
anelastic approximation
\be
\label{equation}
\nabla \times \frac{\nabla p'}{\rho_{\rm o}(r)}= (0,{\omega_\theta}^*,{\omega_\phi}^*)
\ee
where ${\omega_i}^*$ is related to  vorticity production.
In the  anelastic model there is zero contribution to radial vorticity production from the pressure gradient  term.
%Hence, it may require full compressibilty to reproduce smaller scale features of differential
%rotation such as the Near Surface Shear Layer.
It is unclear whether solar differential rotation is being driven primarily
from above or below the convection zone. If it is being driven
by  surface cooling,
as suggested by \cite{CossetteJ16} and \cite{KapylaP17}, then it might require a  fully compressible model to reproduce features such as the near surface shear layer.
\newpage

%\begin{thebibliography}{99}
%\end{thebibliography}

%%%%%%%%%%%%%%%%%%%%%%%%%%%%%%%%%%%%%%%%%%%%%%%%%%

%%%%%%%%%%%%%%%%% APPENDICES %%%%%%%%%%%%%%%%%%%%%

%%%%%%%%%%%%%%%%%%%%%%%%%%%%%%%%%%%%%%%%%%%%%%%%%%

% Don't change these lines
\bsp	% typesetting comment
\label{lastpage}
\end{document}